\documentclass[journal]{IEEEtran}

\usepackage{graphicx}

\ifCLASSINFOpdf
\else
\fi
\usepackage{amsmath}
\usepackage{algorithmic}
\usepackage{multirow}

\DeclareUnicodeCharacter{2212}{-}

\begin{document}
%
\title{Low Power Neuromorphic EMG Gesture Classification}
%
%
%

\author{Sai~Sukruth~Bezugam,
Ahmed~Shaban, Manan~Suri
\thanks{S.S.B, A.S, M.S are with the Department
of Electrical Engineering department, Indian Institute of Technology Delhi, e-mail: manansuri@ee.iitd.ac.in.}
}

%
%

\markboth{A version of this study is under review}%
{S.S.Bezugam \MakeLowercase{\textit{et al.}}: Low Power Neuromorphic EMG Gesture Classification}
%



\maketitle

\begin{abstract}
EMG (Electromyograph) signal based gesture recognition can prove vital for applications such as smart wearables and bio-medical neuro-prosthetic control. Spiking Neural Networks (SNNs) are promising for low-power, real-time EMG gesture recognition, owing to their inherent spike/event driven spatio-temporal dynamics [1], [3]–[5]. In literature, there are limited demonstrations of neuromorphic hardware implementation (at full chip/board/system scale) for EMG gesture classification. Moreover, most literature attempts exploit primitive SNNs based on LIF (Leaky Integrate \& Fire) neurons. In this work, we address the aforementioned gaps with following key contributions: (1) Low-power, high accuracy demonstration of EMG-signal based gesture recognition using neuromorphic Recurrent Spiking Neural Networks (RSNN). In particular, we propose a multi-time scale recurrent neuromorphic system based on special double-exponential adaptive threshold (DEXAT) [2] neurons. Our network achieves state-of-the-art classification accuracy (90\%) while using $\sim$ 53\% lesser neurons than best reported prior art on Roshambo EMG dataset [5]. (2) A new multi-channel spike encoder scheme for efficient processing of real-valued EMG data on neuromorphic systems. (3) Unique multi-compartment methodology to implement complex adaptive neurons on Intel’s dedicated neuromorphic Loihi chip is shown. (4) RSNN implementation on Loihi (Nahuku 32) achieves significant energy/latency benefits of $\sim$ 983X/19X compared to GPU for batch size = 50. 
\end{abstract}


%
\IEEEpeerreviewmaketitle

\section{Introduction}
Roshambo game EMG dataset [1] includes three hand gestures (Rock, Paper, Scissors) and is generated using a commercial Myo armband sensor (see Fig.1 (a)). For each gesture, EMG signal voltage data output is extracted from 8 channels (see Fig. 1 (b)). We convert real-valued EMG voltage data to spike domain using a new spike encoder for multi-channel parallel encoding. Encoding neurons per EMG channel are optimally chosen to attain high classification accuracy. Fig. 1(c) illustrates the modified spike encoding scheme, that exploits ONSET, OFFSET and Touch neurons for the input layer of the network. Each neuron is assigned with a threshold taken from a linearly spaced EMG voltage range (lying between -2 V to +2 V). An ONSET neuron spikes when there is a transition from a lower value to a higher value while an OFFSET neuron spikes for vice versa. If there is no transition at the highest threshold value then a Touch neuron spikes. Fig. 1 (d) illustrates spike trains generated for single channel EMG recording with 9 encoding neurons (4 ONSET, 4 OFFSET and 1 touch). 

\begin{figure}
    \centering
    \includegraphics[width=\linewidth]{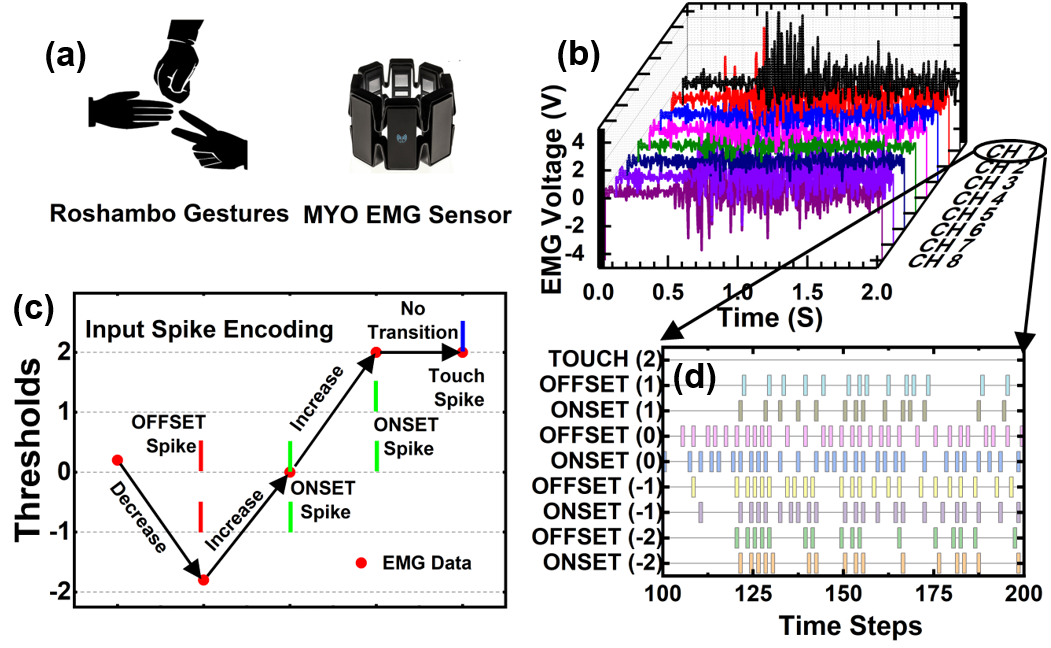}
    \caption{Illustration of (a) Hand gestures (Rock, paper, scissors), myo EMG sensor. (b) 8-channel voltage data readout for class ’Rock’. (c) Input spike encoding methodology. (d) shows generated spike trains through ONSET-OFFSET scheme for 100 time steps for EMG channel-1 (between 0.5 s to 1.0 s).}
    \label{fig:my_label}
\end{figure}

We propose a hybrid RSNN that includes both LIF and DEXAT neurons (Fig. 2(a)). DEXAT neuron model combines the advantages of short and long adaptation time constants (Fig. 2(b)) for spatio-temporal classification on neuromorphic systems [2]. DEXAT ensures local fine tuning of synaptic weights. Input spike encoding layer is connected to a hidden recurrent layer, consisting of ‘m’ LIF and ‘n’ DEXAT neurons. The recurrent layer is connected to an output classification layer consisting of 3 linear readout neurons. Each readout neuron represents one of the three gestures. The network was trained using back propagation through time (BPTT) algorithm. Highest classification accuracy network graph is obtained for 450 epochs with learning rate = 1e$−$2. DEXAT neuron parameters used were $\tau_{a}$1 = 21 $\delta$t, $\tau_{a}$2 = 400 $\delta$t, $\tau_{m}$ = 20 $\delta$t, $\beta$1 = $\beta$2 = 1 (Fig. 2(b)). We choose optimum hidden layer dimensions for the system by sweeping number of hidden neurons (Fig. 2(c)). Network accuracy improves with increasing number of hidden neurons, reaching a peak value for m = 50 LIF and n = 100 DEXAT neurons. Fig. 2 (d) shows train and test accuracy evolution for optimum RSNN size. Best accuracy of 90\% (see Fig. 3) with a standard deviation of 0.8 \% was obtained over 4 iterations. 

\begin{figure}
    \centering
    \includegraphics[width=\linewidth]{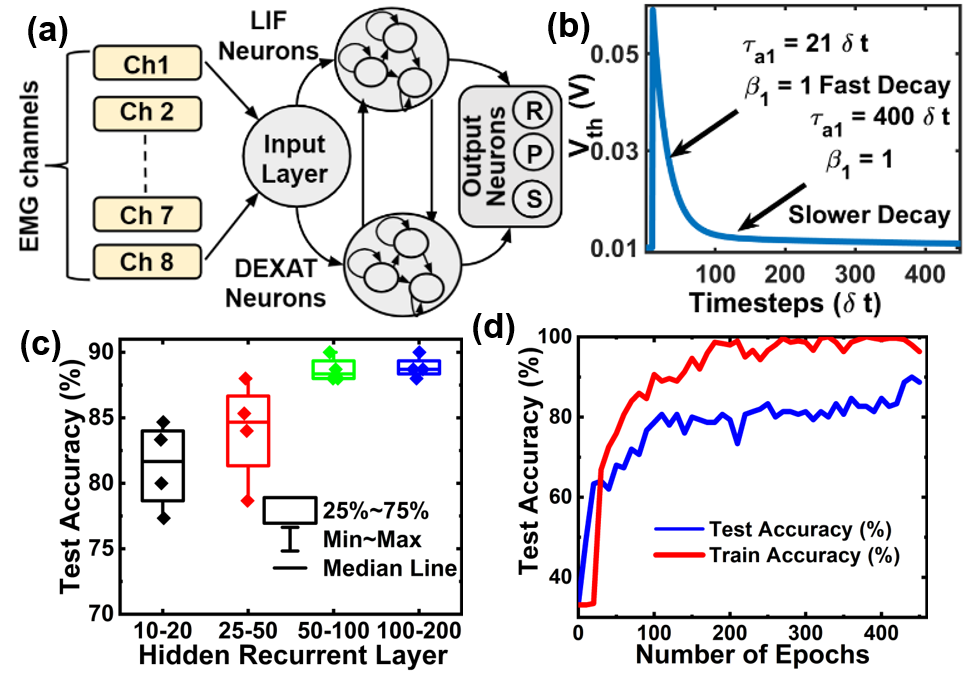}
    \caption{Caption(a) Proposed DEXAT based RSNN for 8-channel EMG gesture classification. (b) DEXAT adaptive threshold behaviour. (c) Box plot (4 runs) showing test accuracy for varying network sizes. (d) Train and test accuracy vs epochs for best accuracy RSNN (50 LIF + 100 DEXAT neurons).}
    \label{fig:my_label}
\end{figure}

Fig. 3(a) shows proposed three compartment design used to realize DEXAT neuron on Intel’s Loihi neuromorphic chip. Loihi supports in-built compartment based LIF neurons with a fixed firing threshold. In our proposed design, input compartment ‘N1’ receives incoming spikes due to which membrane potential rises and a spike is generated once threshold (Vth) is crossed. Following the spike event, membrane potential in primary compartment is reset. Output spike is scaled by negative factors $\gamma$1 and $\gamma$2, using inhibitory synapses (Fig. 3(a)), and fed as input to the two secondary neuron compartments ‘N2’ and ‘N3’. Consequently, membrane potential of ‘N2’ and ‘N3’ rise and decay exponentially following the dynamics of a LIF neuron. Membrane potentials of ‘N2’ and ‘N3’ are exploited to realize the two individual exponential time constants $\tau_{a}$1 and $\tau_{a}$2 of DEXAT. Sum of the membrane potentials of secondary compartments is subtracted from the membrane potential of primary compartment. As a result, after every spike event the membrane potential of the primary compartment decreases in negative direction and the difference Vmem $−$ Vth widens resulting in desired adaptive behaviour. We empirically deduce system level parameter mapping between software DEXAT neuron and its equivalent Loihi hardware implementation (Fig. 3(b)). Inhibitory synaptic weight range in Loihi is limited to [$−$28 , 0], thus ’$\gamma$1’ and ’$\gamma$2’ must be less than 256. Valid range of acceptable parameter values ($\gamma$, $\tau$a, $\beta$) is shown in Fig. 4(b). In order to realize a full classification system on Loihi, we first train the quantization aware RSNN network offline on GPU using tensorflow and then port 8-bit quantized weights to the network on Loihi. LIF and DEXAT neurons are configured in Loihi neuromorphic cores (NC) through compartment prototypes while synapses are configured through connection prototypes. Code is written in NxSDK using NxNet api and executed on Nahuku 32 board (consisting of 32 Loihi chips). Entire 50-100 size RSNN efficiently fits on a single neuromorphic core of Loihi. 

\begin{figure}
    \centering
    \includegraphics[width=\linewidth]{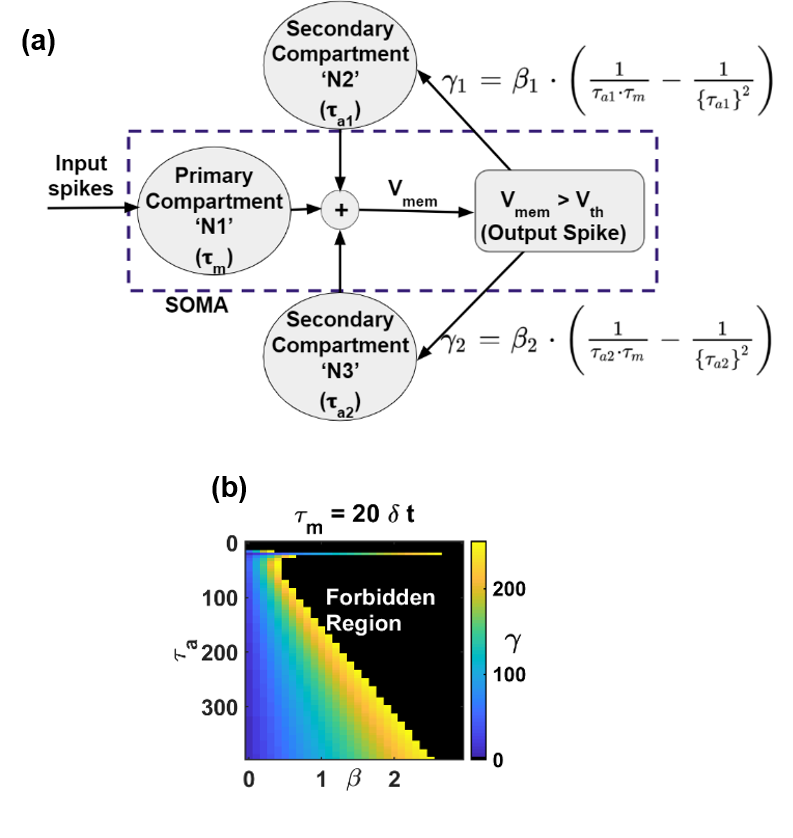}
    \caption{(a) Proposed methodology to implement complex multi-timescale adaptive neurons on Loihi (DEXAT using 3 compartments). (b) System level empirical parameter mapping for hardware DEXAT on Loihi. (Black represents out of hardware range)}
    \label{fig:my_label}
\end{figure}

\begin{table}[]
\caption{Comparison with existing literature on Roshambo dataset}
\label{tab2}
\centering

\begin{tabular}{|c|c|c|c|c|}
\hline
\textbf{Ref}                       & \textbf{Network}                                                                      & \multicolumn{1}{l|}{\textbf{Implementation}}                                      & \textbf{\begin{tabular}[c]{@{}c@{}}Number \\ of \\ Neurons\end{tabular}} & \textbf{\begin{tabular}[c]{@{}c@{}} S/W Test \\ Accuracy \\ (\%)\end{tabular}} \\ \hline
\multirow{3}{*}{AICAS'20}          & \begin{tabular}[c]{@{}c@{}}LSM +\\ SVM\end{tabular}                                   & \multirow{3}{*}{\begin{tabular}[c]{@{}c@{}}S/W \\ \& \\ DynapSE\end{tabular}}     & 320                                                                      & 73.2                                                                      \\ \cline{2-2} \cline{4-5} 
                                   & \begin{tabular}[c]{@{}c@{}}LSM  + \\ Spike rate \\ distance\end{tabular}              &                                                                                   & 320                                                                      & 77.5                                                                      \\ \cline{2-2} \cline{4-5} 
                                   & \begin{tabular}[c]{@{}c@{}}LSM  + \\ Trace STDP\end{tabular}                          &                                                                                   & 320                                                                      & 61.4                                                                      \\ \hline
\multirow{3}{*}{BIOCAS'19}          & \begin{tabular}[c]{@{}c@{}}Logistic \\ regression\end{tabular}                        & Only S/W                                                                               &   -                                                                       & 81.0                                                                      \\ \cline{2-5} 
                                   & SVM                                                                                   & Only S/W                                                                               &  -                                                                        & 84.0                                                                      \\ \cline{2-5} 
                                   & \begin{tabular}[c]{@{}c@{}}Feedforwad\\ SNN\end{tabular}                              & \begin{tabular}[c]{@{}c@{}}S/W \& \\ DynapSE\end{tabular}                         & 192                                                                      & 74.0                                                                      \\ \hline
\multirow{2}{*}{JETCAS'20}          & \begin{tabular}[c]{@{}c@{}}Reservoir + \\ spike-rate \\ distance\end{tabular}         & \begin{tabular}[c]{@{}c@{}}S/W \&\\ DynapSE\end{tabular}                          & 320                                                                      & 85.3                                                                      \\ \cline{2-5} 
                                   & \begin{tabular}[c]{@{}c@{}}Reservoir + \\ SVM\end{tabular}                            & \begin{tabular}[c]{@{}c@{}}S/W \&\\ DynapSE\end{tabular}                          & 320                                                                      & 75.0                                                                      \\ \hline
\multirow{4}{*}{ICONS'21}          & \begin{tabular}[c]{@{}c@{}}Reservoir +\\ LDA\end{tabular}                             & \multirow{4}{*}{Only S/W}                                                         & 320                                                                      & 76.5                                                                      \\ \cline{2-2} \cline{4-5} 
                                   & \begin{tabular}[c]{@{}c@{}}Reservoir +\\ SVM\end{tabular}                             &                                                                                   & 320                                                                      & 76.5                                                                      \\ \cline{2-2} \cline{4-5} 
                                   & \begin{tabular}[c]{@{}c@{}}Reservoir \\ (Plastic) + \\ LDA\end{tabular}              &                                                                                   & 320                                                                      & 83.1                                                                      \\ \cline{2-2} \cline{4-5} 
                                   & \begin{tabular}[c]{@{}c@{}}Reservoir \\ (Plastic) +\\ SVM\end{tabular}               &                                                                                   & 320                                                                      & 88.0                                                                      \\ \hline
\multirow{3}{*}{\textbf{Proposed}} & \multirow{3}{*}{\textbf{\begin{tabular}[c]{@{}c@{}}DEXAT \\ based RSNN\end{tabular}}} & \multirow{3}{*}{\textbf{\begin{tabular}[c]{@{}c@{}}S/W \& \\ Loihi\end{tabular}}} & \textbf{30}                                                              & \textbf{84.7}                                                             \\ \cline{4-5} 
                                   &                                                                                       &                                                                                   & \textbf{75}                                                              & \textbf{88.0}                                                             \\ \cline{4-5} 
                                   &                                                                                       &                                                                                   & \textbf{150}                                                             & \textbf{90.0}                                                             \\ \hline
\end{tabular}
\end{table}

We performed system level performance profiling across 3 hardware platforms (CPU, GPU, and Loihi) for different batch sizes (1 and 50) (Fig. 4). ‘Nvidia-smi’ api and ‘Intel power gadget’ were used to profile power for GPU and CPU variants of the RSNN respectively. While in-built energy probes of Nahuku were used for power measurements of Loihi. For all cases, we ensure that only RSNN network is run on the host computer during power profiling. Python time library was used to characterize inference latency. Hardware implementation on Loihi shows promising results with very low energy dissipation/per inference (0.37 mJ) owing to its asynchronous architecture and fine granularity intrinsic support for spiking neuron computations [6]. Inference dissipation primarily constitutes three types of operations i.e., spike encoding of EMG data, neuronal activity (DEXAT/LIF) and synaptic operations of the recurrent layers. 

\begin{figure}
    \centering
    \includegraphics[width=\linewidth]{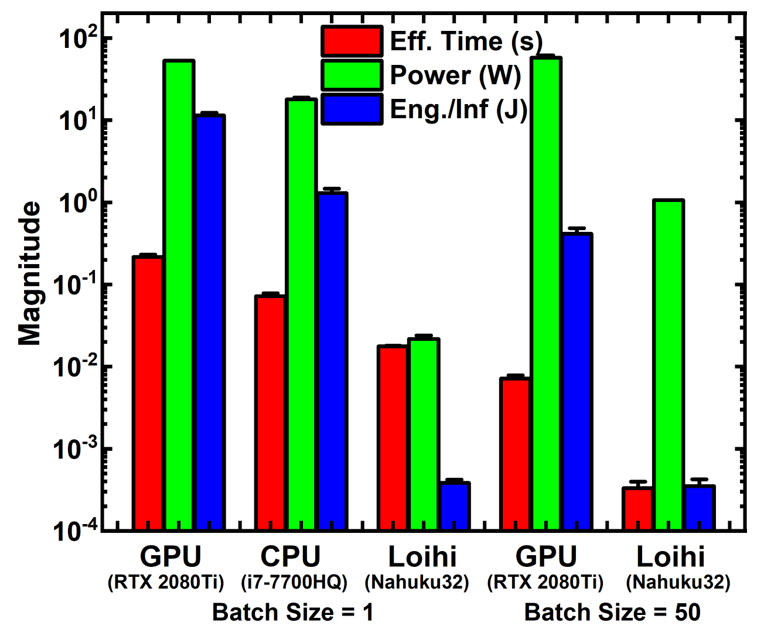}
    \caption{Performance benchmarking w.r.t GPU,CPU with varying batch-sizes. Error-bar over 5 runs is included on each datapoint.}
    \label{fig:my_label}
\end{figure}

Operation specific benchmarking is shown in Fig. 5. Loihi delivers significant energy/latency gains of $\sim$ 3363X/4X and $\sim$ 29683X/12X w.r.t CPU and GPU respectively (for batch size = 1). Batch-size = 1 is representative of real-time or live online processing use-case, while batch-size > 1 corresponds to stored/buffered data processing. Batch size increment on Loihi is limited by Lakemont x86 cores that store real valued data and perform spike encoding before pushing encoded spike-data on the neuromorphic ASIC SNN. We emulated an artificial batch size = 50, for Loihi by pre-generating the encodings and duplicating 50 RSNNs on a single Loihi chip. Clearly GPU offers latency $\sim$ 10X gain for multi-batch encoding (Fig.5), however Loihi’s 5X latency gains in neuronal computations makes its overall inference $\sim$ 19X faster. 

\begin{figure}
    \centering
    \includegraphics[width=\linewidth]{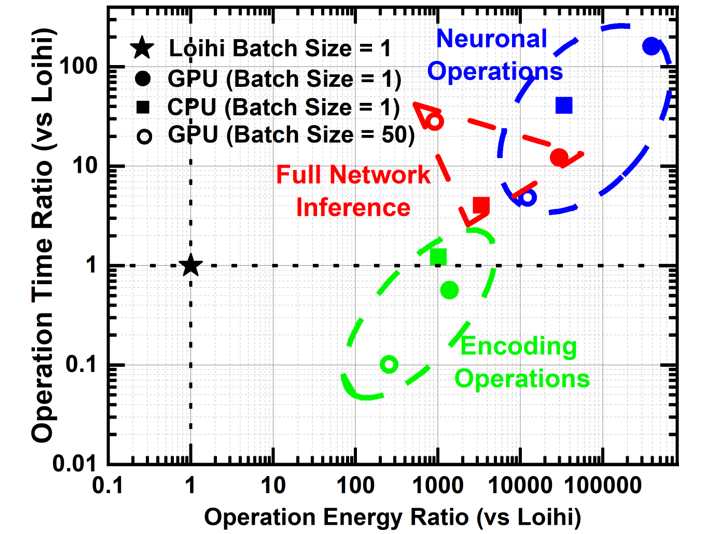}
    \caption{Energy- and latency- ratio w.r,t Loihi (batch size = 1) for proposed network for CPU and GPU implementations. Green, blue and red colors denote encoding, neuronal and full network operations respectively. For hidden layer spiking neuronal operations Loihi provides drastic energy gains w.r.t. CPU/GPU.}
    \label{fig:my_label}
\end{figure}

\section*{Acknowledgment}

Authors would like to thank Intel Labs for providing access to Loihi/Nahuku neuromorphic platforms and CYRAN AI Solutions for partial support.

\ifCLASSOPTIONcaptionsoff
  \newpage
\fi



%

%

\end{document}